\documentclass{article}
\usepackage{spconf,amsmath,amsfonts, amssymb, graphicx,amsthm,epsfig}
\usepackage{bm}  
\usepackage{color}
\usepackage{amssymb}
\usepackage{dsfont}
\usepackage{mathtools}
\usepackage{array,multirow}

\newcommand{\bxi}{\bm{\xi}}
\newcommand{\bpsi}{\bm{\psi}}
\newcommand{\bmu}{\bm{\mu}}
\newcommand{\bmeta}{\bm{\eta}}
\newcommand{\btheta}{\bm{\theta}}
\newcommand{\e}{\mathbb{E}}
\newcommand{\f}{\mathcal{F}}
\newcommand{\bmf}{\boldsymbol{\mathcal{F}}}

\newcommand{\dkl}{D_{\text{KL}}}
\newcommand{\dtv}{D_{\text{TV}}}
\newcommand{\mcl}{\text{col}}
\DeclareMathOperator{\T}{\mathsf{T}}

\newtheorem{theorem}{Theorem}
\newtheorem{corollary}{Corollary}
\newtheorem{assumption}{Assumption}

\let\OLDthebibliography\thebibliography
\renewcommand\thebibliography[1]{
  \OLDthebibliography{#1}
  \setlength{\parskip}{0pt}
  \setlength{\itemsep}{0pt plus 0.3ex}
}

\begin{document}\sloppy

\def\x{{\mathbf x}}
\def\L{{\cal L}}

\title{HIDDEN MARKOV MODELING OVER GRAPHS}
%
\name{Mert Kayaalp$^{\star}$, Virginia Bordignon$^{\star}$, Stefan Vlaski$^{\dagger}$, and Ali H. Sayed$^{\star}$ \thanks{This work was supported in part by SNSF grant 205121-184999. Emails: mert.kayaalp@epfl.ch, virginia.bordignon@epfl.ch , s.vlaski@imperial.ac.uk , ali.sayed@epfl.ch .}}
\address{$^{\star}$\'{E}cole Polytechnique F\'{e}d\'{e}rale de Lausanne (EPFL)\\ $^{\dagger}$ Imperial College London}

\maketitle

\begin{abstract}
This work proposes a multi-agent filtering algorithm over graphs for finite-state hidden Markov models (HMMs), which can be used for sequential state estimation or for tracking opinion formation over dynamic social networks. We show that the difference from the optimal centralized Bayesian solution is asymptotically bounded for geometrically ergodic transition models. Experiments illustrate the theoretical findings and in particular, demonstrate the superior performance of the proposed algorithm compared to a state-of-the-art social learning algorithm. 
\end{abstract}
\section{Introduction and Related Work}
\label{sec:intro}

We consider a network of agents observing data that are emitted by some (hidden or) latent state of a dynamic Markov system. The goal is to cooperatively infer and track the time-varying state. The state can be any quantity of interest, e.g., location of a moving object, concentration of air pollutants, or whether it is a sunny or rainy day. The model is general enough and can be used for many applications such as source localization, environmental monitoring, target tracking, navigation, and analyzing opinion formation in social networks. 

There exist several works addressing the decentralized multi-agent state estimation problem. One body of work considers distributed Kalman filters \cite{olfati_2007,khan_2008,cattivelli_2010,talebi2021}. These papers assume linear state dynamics and observations, whereas this work does not make these assumptions. For non-linear system models, some works use a Bayesian framework \cite{hlinka_2012, battistelli_2014, dedecius_2017, bandyopadhyay_2018}, which we also adopt. It is based on calculating \emph{beliefs}, which are distributions over the set of states. Typically, this involves temporal recursions that consist of computing the new beliefs from the previous beliefs by taking the observed data into account. The Bayesian framework is advantageous in the sense that it uses the complete information of a distribution rather than only some statistics of it such as the mean. Moreover, it enjoys optimality in the minimum mean-square-error and the maximum a-posteriori sense for the single agent mode of operation \cite{krishnamurthy_2016}. 

Among works that examine distributed Bayesian state estimation, the articles \cite{hlinka_2012, battistelli_2014} require multiple rounds of communication between agents for every state change. This might be impractical especially when the state changes rapidly. We will show that it is sufficient to communicate once per iteration to attain a bounded difference from the optimal estimator by means of our proposed solution. The works \cite{hlinka_2012, dedecius_2017} also consider observation likelihood functions that belong to the exponential family of distributions in order to have numerically tractable solutions. In our work, we consider finite dimensional HMMs where the true state/hypothesis can take only a finite set of values. Therefore, we do not need to restrict the distributions to analytically well-behaved families of distributions. Likewise, sub-optimal filters like particle filters \cite{sayed_2018,papa2019} are not necessary in this case. On the other hand, the work \cite{bandyopadhyay_2018} proposes a distributed Bayesian filtering algorithm that does not require multiple consensus steps at every iteration. However, the agents are assumed to use an \emph{effective} likelihood in place of their own likelihood function. Calculating the effective likelihood includes combining the neighbors' effective likelihoods from the previous time instant with their likelihood from that time instant. As the assumptions in the convergence analysis suggest, this would be beneficial only if the likelihood functions are changing slowly over time. This limits the maximum rate of state transition the algorithm can track. In comparison, our algorithm combines the likelihoods with \emph{time-adjusted} priors, in which the neighbors' beliefs at the previous time instant are combined to compute the priors. Consequently, we get bounds (Theorem \ref{th:kl_without_net_assumption}) that hold for a large class of transition models, namely, geometrically ergodic models, which includes rapidly mixing Markov chains. 

Combining the neighbors' beliefs is also useful for modeling opinion formation over social networks. In these models, agents' beliefs/opinions over the set of states/hypotheses are formed based on their local observations and their interactions with the other agents \cite{acemoglu_2011, krishnamurthy_2013, jadbabaie_2012, zhao_2012}. A line of social learning algorithms comprises of two iterative steps. First, each agent revises its belief via a Bayesian update based on new private data. Second, agents aggregate their neighbors' information into their beliefs with a distributed learning algorithm like consensus \cite{jadbabaie_2012,nedic_2017,lalitha_2018,luengo2018} or diffusion \cite{zhao_2012, matta_2020, ntemos_2021}. A common assumption in these works is that the state of nature is fixed. In many practical applications, however, the state is time-varying. Linear transition models are considered in \cite{acemoglu_2008, frongillo_2011, shahrampour_2013}. However, to tackle the possible drifts in the state of nature, an \emph{adaptive social learning} (ASL) strategy was proposed in \cite{bordignon_2021}. This algorithm enables the agents to respond to the changes faster. Nevertheless, it does not take the transition model into account. In many scenarios, the current true state will make some states more likely to occur than others in the future. By exploiting knowledge about the system dynamics, the true state can be tracked better. Indeed, we confirm this statement experimentally by comparing ASL against our proposed algorithm in Section \ref{sec:simulations}. \\

\noindent \textbf{Contributions}. We propose a distributed Bayesian HMM filtering algorithm for detecting a time-varying state in Section \ref{sec:dif_HMM}. The algorithm requires only one round of communication per state change and takes advantage of prior information about state dynamics, which allow tracking rapidly changing states. We examine how close this distributed strategy gets to the optimal centralized solution in Section \ref{sec:main_results}. More specifically, an asymptotic bound on the expected Kullback-Leibler (KL) divergence between the centralized belief and the agent-specific beliefs is established in Theorem \ref{th:kl_without_net_assumption} for geometrically ergodic transition models. Corollary \ref{co:belief_bound} relates this bound to belief values evaluated at the true hypothesis. Simulation results in Section \ref{sec:simulations} support the theoretical results and compare the proposed algorithm with a state-of-the-art algorithm, ASL \cite{bordignon_2021}.

\section{ALGORITHM DESCRIPTION}
\label{sec:alg_description}

A network of \( K \) communicating agents are exchanging beliefs with each other in order to keep track of the underlying state of nature, which is allowed to evolve over time. The belief of agent \( k \in \mathcal{N}\), at time \( i \), is denoted by \( \mu_{k,i} \) and it is a probability simplex. The value of \( \mu_{k,i} (\theta) \) represents the probability that agent \( k \) believes the hypothesis \( \theta \in \Theta \) is the true hypothesis at time \( i \). The true hypothesis at that same time instant is denoted by \( \btheta_i^\circ \in \Theta \), and it is assumed to belong to a finite set of \( H \) hypotheses, \( \Theta = \{0,1,..,H-1\}\). Note that we are using boldface letters to refer to random variables. The transition model \( T \), which is assumed to be known to the agents, is a Markov chain and we use the notation:
\begin{align}
T (\theta_i | \theta_{i-1}) \triangleq \mathbb{P} (\btheta_i^\circ = \theta_i | \btheta_{i-1}^\circ = \theta_{i-1})
\end{align}
At instant \( i \), each agent \( k \) observes a private observation \( \bxi_{k,i}  \) distributed according to the agent-specific likelihood function \( L_k (\bxi_{k,i} | \btheta_i^\circ)\), which is known to agent \( k \). The likelihoods can be probability mass or density functions depending on whether observations are discrete or continuous.
\begin{assumption}\label{as:independence}\cite{hlinka_2012,shahrampour_2013}
The observations are assumed to be independent across agents given the state. Denoting the joint observations by \( \bxi_i \triangleq \{ \bxi_{k,i} \}_{k=1}^K \) and its distribution by \( L^\star (\bxi_i | \btheta_i^\circ) \), we have for all \( \xi_i \triangleq \{ \xi_{k,i} \}_{k=1}^K  \) and \( \theta_i \in \Theta \):
\begin{align}
L^\star (\xi_i | \theta_i) = \prod_{k=1}^K L_k (\xi_{k,i} | \theta_i)
\end{align}
\end{assumption} \qed

Agents can communicate with each other once per iteration. We have the following assumption on the communication topology.
\begin{assumption}\label{as:network_top}
The communication topology underlying the network is a strongly-connected graph \cite{sayed_2014}. This means that the combination matrix \( A \triangleq [a_{\ell k}]\) is a primitive matrix. The coefficient \( a_{\ell k}\) weights the information sent by agent \( \ell \) to \( k \) and is nonzero if, and only if, \( \ell\in\mathcal{N}_k\) (i.e., for every agent \(\ell\) in the neighborhood of \(k\)). The matrix \( A \) is doubly-stochastic and symmetric, i.e., it satisfies:
\begin{align}
A\mathds{1}_{K}=\mathds{1}_{K}, \quad A=A^{\T}
\end{align}
\qed
\end{assumption} 
\noindent We also assume a regularity condition on the likelihood functions.
\begin{assumption}\label{as:likelihood_functions}
For each agent \( k \), log-likelihoods are bounded in absolute value, namely:
\begin{align}
    |\log L_k (\cdot | \cdot)| \leq \alpha
\end{align}
\qed
\end{assumption}
\noindent For instance, this assumption is satisfied for truncated Gaussian likelihoods.
\subsection{Optimal Centralized Belief Recursion}\label{sec:centralized_HMM}
Given the observation history of all agents \( \bmf_i \triangleq \{\bxi_j \}_{j=1}^i \), the posterior distribution of the true hypothesis at time \( i \) is denoted by:
\begin{align}
\bmu_i^\star (\theta_i) \triangleq \mathbb{P} (\btheta_i^\circ = \theta_i | \bmf_i)
\end{align}
This posterior satisfies the optimal Bayesian filtering recursion \cite{krishnamurthy_2016}:
\begin{align}\label{eq:centralized_posterior}
 \bmu_i^\star (\theta_i) = \frac{L^\star(\bxi_i|\theta_i) \bmeta_i^\star (\theta_i)}{ \sum_{\theta_{i}^\prime} L^\star(\bxi_i|\theta_i^\prime)\bmeta_i^\star (\theta_i^\prime)}
\end{align}
where \( \bmeta_i^\star (\theta_i) \) is the time-adjusted prior at time instant \( i \) and given by:
\begin{align}\label{eq:ck_cent}
\bmeta_i^\star (\theta_i) \triangleq \sum_{\theta_{i-1}} T(\theta_{i}| \theta_{i-1}) \bmu_{i-1}^\star(\theta_{i-1})
\end{align}
In the sequel, we study how close the beliefs generated by the proposed distributed algorithm get to the above centralized posterior \eqref{eq:centralized_posterior}.
\subsection{Diffusion HMM Strategy}\label{sec:dif_HMM}
Agents across the network need to update their beliefs based on their local streaming observations, as well as exchange their beliefs with each other in order to track the true state in the face of stochastic and dynamic conditions. To do so, we propose a \emph{social} HMM filtering algorithm that is based on the diffusion strategy for cooperation over networks \cite{sayed_2014}. At every time instant \( i \), agents first revise their belief at \( i-1 \) via the Chapman-Kolmogorov equation \cite{krishnamurthy_2016}:
\begin{align}\label{eq:dif_evolve_step}
\bmeta_{k,i} (\theta_{i}) &= \sum_{\theta_{i-1}} T(\theta_{i}| \theta_{i-1}) \bmu_{k,i-1}(\theta_{i-1}) \qquad \text{(Evolve)}
\end{align}
Then, each agent \( k \) forms an intermediate belief \emph{locally} by a \emph{\( \gamma\)-scaled} Bayesian update based on the received data:
\begin{align}\label{eq:dif_adapt_step}
 \bpsi_{k,i} (\theta_{i}) &= \frac{(L_k(\bxi_{k,i} | \theta_{i}))^{\gamma}\bmeta_{k,i} (\theta_{i})}{\sum_{\theta_{i}^\prime}(L_k(\bxi_{k,i} | \theta_{i}^\prime))^{\gamma}\bmeta_{k,i} (\theta_{i}^\prime)} \qquad \text{(Adapt)}
\end{align}
where \( \gamma > 0 \) is a step-size that scales the newly arrived data against prior information. Finally, agents combine the intermediate beliefs of their neighbors into their updated belief:
\begin{align}\label{eq:dif_combine_step}
   \bmu_{k,i}(\theta_{i}) &= \frac{ \text{exp}\{\sum_{\ell \in \mathcal{N}_k} a_{\ell k} \log \bpsi_{\ell,i} (\theta_{i}) \}} {\sum_{\theta_{i}^\prime} \text{exp}\{{\sum_{\ell \in \mathcal{N}_k} a_{\ell k} \log \bpsi_{\ell,i} (\theta_{i}^\prime)}\}}   \quad \text{(Combine)}
\end{align}
Repeatedly exchanging and fusing beliefs will allow the local information to diffuse throughout the network.

Note that the proposed algorithm recovers standard log-linear social learning algorithms \cite{nedic_2017, lalitha_2018, matta_2020, ntemos_2021} when agents perform local Bayesian updates with \( \gamma=1 \) and the true hypothesis is \emph{fixed}:
\[ T(\theta_i | \theta_{i-1}) = \begin{cases} 
      1, & \theta_i=\theta_{i-1} \\
      0, & \theta_i\neq\theta_{i-1}
   \end{cases}
\]
Moreover, the agents' beliefs will match the optimal centralized belief if \( a_{\ell k} = 1/K \) \(\forall \ell,k\in\mathcal{N}\) (i.e., when the  network is fully-connected), all priors are equal (\(\mu_0^\star = \mu_{k,0}\) \(\forall k \in \mathcal{N}\)) and \( \gamma = K \). These two special cases motivate us to use a general step-size \( \gamma > 0\).
\section{MAIN RESULTS}\label{sec:main_results}
To avoid discarding any hypothesis in the beginning, we have an assumption on the initial values of the beliefs.
\begin{assumption}\label{as:positive_initial_beliefs}\cite{bandyopadhyay_2018,nedic_2017}
We assume that all initial beliefs are strictly positive at every hypothesis, i.e., for each hypothesis \( \theta \in \Theta \) and for each agent \( k \), \( \mu_{k,0} (\theta) > 0 , \mu_0^\star(\theta) > 0\).
\qed
\end{assumption}
We can assess the performance of the algorithm \eqref{eq:dif_evolve_step}--\eqref{eq:dif_combine_step} relative to the optimal centralized solution \eqref{eq:centralized_posterior} by considering time-varying risks of the form:
\begin{align}\label{eq:filter_risk}
J_i(\bmu_{k,i})&\triangleq\e_{\f_i} \dkl(\bmu_i^\star || \bmu_{k,i}) 
\end{align}
and
\begin{align}\label{eq:prior_risk}
\widetilde{J}_i(\bmeta_{k,i})&\triangleq\e_{\f_{i-1}} \dkl(\bmeta_i^\star || \bmeta_{k,i}) 
\end{align}
for each agent \( k \), where \( \dkl(\cdot || \cdot) \) represents the KL divergence, and \(\e_{\f_i}\) denotes the expectation over \( \f_i \) with respect to \( \mathbb{P} \). The risk in \eqref{eq:filter_risk} measures the disagreement between agent \( k \) and the centralized belief for the true state at time instant \( i \) after the observations are emitted from that state. In comparison, the risk in \eqref{eq:prior_risk} is the disagreement before the observations are emitted. In other words, \( J_i \) is the \emph{posterior} divergence while \( \widetilde{J}_i \) is the divergence of time-adjusted \emph{priors}. 

The effect of the transition model will arise via the Dobrushin coefficient \( \kappa(T)\in [0,1] \) defined as follows \cite{dobrushin_1956,polyanskiy_2017,krishnamurthy_2016}:
\begin{align}
\kappa (T) = \sup_{\theta,\theta^\prime \in \Theta} \dtv \Big (T(\cdot |\theta) , T(\cdot |\theta^\prime) \Big)
\end{align}
where \( \dtv(\cdot , \cdot) \) represents the total variation distance. For instance, for a binary symmetric channel:
\[ T(\theta_i | \theta_{i-1}) = \begin{cases} 
      1-\delta, & \theta_i=\theta_{i-1} \\
      \delta, & \theta_i\neq\theta_{i-1}
   \end{cases}
\]
we have \( \kappa(T) = | 1-2\delta |\). In general, the closer the Dobrushin coefficient is to zero, the faster the forgetting of the initial conditions will be.

For ease of notation, we define the column vector consisting of marginal likelihoods over hypotheses and agents of a joint observation \( \xi_j \) as:
\begin{align}
\mathcal{L}_{\xi_j} \triangleq \mcl \Big \{ \mcl \Big \{   \log L_\ell(\xi_{\ell,j} | \theta_j)  \Big\}_{\theta_j=1}^H \Big \}_{\ell=1}^K
\end{align}
We first establish that the risk functions are asymptotically bounded for each agent \( k \).
\begin{theorem}\label{th:kl_without_net_assumption}
Under Assumptions \ref{as:independence}, \ref{as:network_top}, \ref{as:likelihood_functions}, \ref{as:positive_initial_beliefs} and geometrically ergodic state transition models \cite{krishnamurthy_2016}, i.e., \( \kappa(T) < 1 \), for each agent \( k \) the risks are asymptotically bounded, namely:
\begin{align}\label{eq:th1_risk}
\limsup_{i \rightarrow \infty} J_i(\bmu_{k,i}) \leq  \frac{2K \gamma  \lambda}{1-\kappa (T)}   \ \sup_{t>0}\e_{\xi_t} \| \bm{\mathcal{L}}_{\xi_t}\|_{\infty}
\end{align}
and
\begin{align}\label{eq:th1_prior}
\limsup_{i \rightarrow \infty} \widetilde{J}_i(\bmeta_{k,i}) \leq  \frac{2\kappa(T)K \gamma  \lambda}{1-\kappa (T)}   \ \sup_{t>0}\e_{\xi_t} \| \bm{\mathcal{L}}_{\xi_t}\|_{\infty}
\end{align}
where \( \lambda \triangleq \max \{ |1-\frac{K}{\gamma}|, \rho_2 \} \), and \( \rho_2 \) is the absolute value of the second largest magnitude eigenvalue of \( A \). 
\end{theorem}
\begin{proof}
Omitted due to space limitations. 
\end{proof}
Note that if the transition Markov chain is stationary, we have, for any \( t^\prime > 0 \):
\begin{align}
    \sup_{t>0}\e_{\xi_t} \| \bm{\mathcal{L}}_{\xi_t}\|_{\infty} = \e_{\xi_{t^\prime}} \| \bm{\mathcal{L}}_{\xi_{t^\prime}}\|_{\infty}
\end{align}
The bounds are tight in the sense that they are equal to zero as expected when the centralized solution is matched with \( a_{\ell k} = 1/K \) \(\forall \ell,k\in\mathcal{N}\) (\( \rho_2 = 0 \)), equal priors (\(\mu_0^\star = \mu_{k,0}\)  \(\forall k \in \mathcal{N}\)), and \( \gamma = K \) . In fact, the bounds \eqref{eq:th1_risk} and \eqref{eq:th1_prior} do not depend on the initial beliefs as long as they satisfy Assumption \ref{as:positive_initial_beliefs}. Geometric ergodicity is sufficient to forget the initial conditions. Specifically, for \( \rho_2 = 0 \) and \( \gamma = K \), the asymptotic risk is zero which means that the filter is \emph{stable}.

Furthermore, for \( \gamma \to K \), the risks are proportional to the mixing rate of the graph, \( \rho_2 \). This factor underlines the benefit of cooperation. More connected graphs, with smaller \( \rho_2 \), will track the centralized solution better whereas sparse networks or non-cooperative agents will have higher deviation from the optimal. The bounds are also proportional to the network size. The disagreement between the agents and the optimal centralized solution, which has access to all data of agents, increases with the number of agents.

Observe that if the transition Markov chain is mixing very fast such that \( \kappa (T) = 0 \), the bound in \eqref{eq:th1_prior} goes to zero. This is expected because the transition model, in equations \eqref{eq:ck_cent} and \eqref{eq:dif_evolve_step}, will output the same distributions and the divergence between them will vanish. Therefore, \eqref{eq:th1_prior} captures the effect of ergodicity accurately. However, the bounds are still not tight enough in the sense that the effect of observations is not sufficiently reflected. For instance, if the state of nature is fixed, we have \( \kappa (T) = 1 \), which is not a geometrically ergodic model. However, from the standard social learning literature \cite{jadbabaie_2012, zhao_2012,nedic_2017,lalitha_2018}, we know that true state can be learned. Bounds that address both ergodicity and informativeness of the measurements is an interesting future work.

Now assume that the risks are asymptotically bounded with \( \limsup_{i \rightarrow \infty} J_i(\bmu_{k,i}) \leq B \) , as suggested by Theorem \ref{th:kl_without_net_assumption}. Next, we relate this upper bound to belief values at the true hypothesis.
\begin{corollary}\label{co:belief_bound}
Define the variance:
\begin{align}\label{eq:variance}
&\text{Var} \Big (\log \frac{\bmu_i^\star (\btheta_i^\circ)}{\bmu_{k,i} (\btheta_i^\circ)} \Big ) \notag \\ &\triangleq \e_{\f_i,\theta_i^\circ} \Big |\log \frac{\bmu_i^\star (\btheta_i^\circ)}{\bmu_{k,i} (\btheta_i^\circ)}- \e_{\f_i,\theta_i^\circ}\Big [\log \frac{\bmu_i^\star (\btheta_i^\circ)}{\bmu_{k,i} (\btheta_i^\circ)} \Big ]   \Big |^2
\end{align}
Then, with probability at least: 
\begin{align}
p \triangleq 1 -  \frac{\text{Var} \Big (\log \frac{\bmu_i^\star (\btheta_i^\circ)}{\bmu_{k,i} (\btheta_i^\circ)} \Big )}{\epsilon^2}
\end{align}
where \( \epsilon \) is an arbitrary positive constant such that \( p \in (0,1] \), the agent and centralized beliefs evaluated at the true hypothesis \( \theta_i^\circ \) satisfy the following relation as \( i\rightarrow \infty \):
\begin{align}
\mu_{k,i} (\theta_i^\circ) &\geq  \mu_i^\star (\theta_i^\circ) \exp \Big \{-\epsilon-B \Big \}
\end{align} 
\end{corollary}
\begin{proof}
Omitted due to space limitations. 
\end{proof}
This is a guarantee on the beliefs evaluated at the true hypothesis. For small risks \eqref{eq:filter_risk} and variances \eqref{eq:variance}, agents will assign a belief value which is close to the value assigned by the optimal belief.

\section{Simulation Results}\label{sec:simulations}
\begin{figure}[ht]
	\centering
	\includegraphics[width=3in]{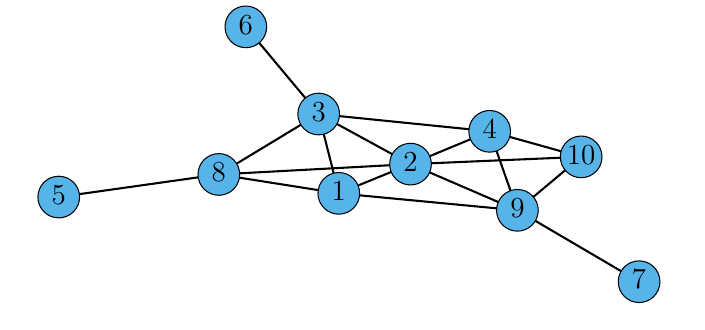}
	\caption{Network diagram.}
	\label{fig:net}
\end{figure}

We consider first a $10-$agent network whose topology is displayed in Fig.~\ref{fig:net}. The combination matrix is given by the Metropolis rule \cite{  metropolis_1953,sayed_2014}, resulting in a doubly-stochastic and symmetric matrix with $\rho_2 = 0.86$. 

The network would like to track the true state of nature from a set of two hypotheses,  $\Theta=\{0,1\}$. We assume that all agents possess the same family of truncated Gaussian likelihoods :
\[ L_k(\xi|\theta)  = \begin{dcases} 
       \frac{1}{Z_{\theta}}\frac{1}{\sqrt{2\pi}}\exp\left\{-\frac{(\xi - (\theta +1))^2}{2}\right\},&-1\leq \xi \leq 2 \\
      \qquad \qquad 0, & \text{otherwise}
   \end{dcases}
\]
for all $k=1,2,\dots, K$, where \( Z_{\theta} \) is the normalization constant:
\begin{align}
    Z_{\theta} \triangleq \int_{-1}^2 \frac{1}{\sqrt{2\pi}}\exp\left\{-\frac{(\xi - (\theta +1))^2}{2}\right\} d\xi
\end{align}
Notice that, in this case, Assumption \ref{as:likelihood_functions} is satisfied. The hidden state is assumed to be a Markovian random variable, whose transition matrix is given by:
\[ T(\theta_i | \theta_{i-1}) = \begin{cases} 
      0.9, & \theta_i=\theta_{i-1} \\
      0.1, & \theta_i\neq\theta_{i-1}
   \end{cases}
\]
which corresponds to the Dobrushin coefficient \( \kappa(T) = 0.8 \). 

\begin{figure}[ht]
	\centering
	\includegraphics[width=3in]{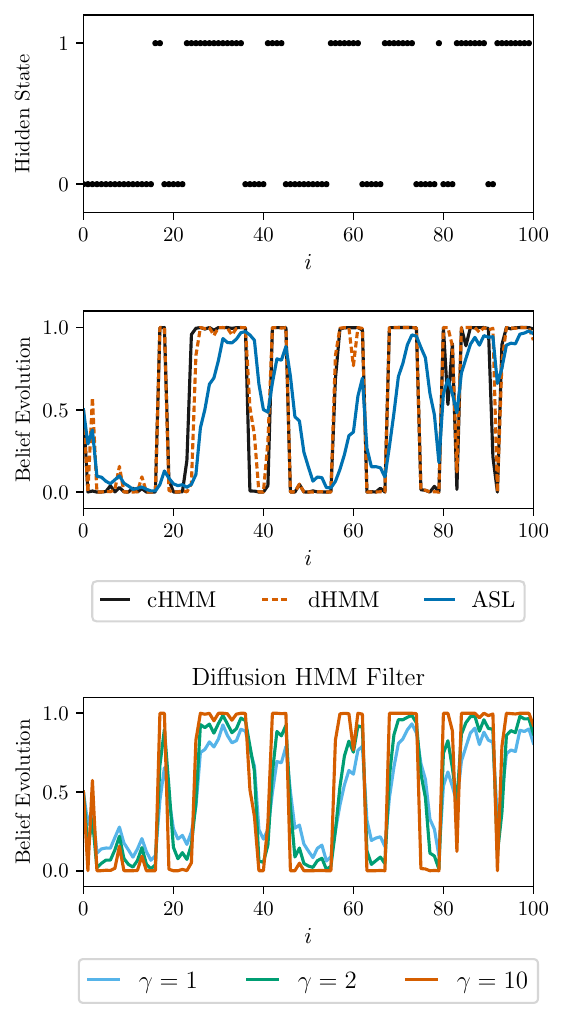}
	\caption{{\em Top panel}: Evolution of the hidden state. {\em Middle panel}: Evolution of beliefs over time for different methods (centralized HMM, diffusion HMM, and ASL). {\em Bottom panel}: Evolution of beliefs over time for the diffusion HMM filter.}
	\label{fig:pl1}
\end{figure}

We first compare the performance of the proposed diffusion HMM filter (dHMM) with the centralized HMM filter (cHMM) for a particular realization of hidden states $\bm{\theta}^{\circ}_{i}$. This comparison is seen in the middle panel of Fig.~\ref{fig:pl1}. Notice that both dHMM, with choice $\gamma=K$, and cHMM behave similarly, which supports Corollary \ref{co:belief_bound}. They show a remarkable capacity of tracking the abrupt changes in the hidden state. In the same figure, we have also included the behavior of the ASL algorithm with a choice of step size $\delta = 0.1$ . ASL does not utilize the transition model knowledge. Hence, it is slower to respond to changes compared to cHMM and dHMM. 

In the bottom panel of Fig.~\ref{fig:pl1}, we explore different choices of $\gamma$ and observe the evolution of beliefs for the distributed HMM filter. We can see that as $\gamma$ approaches $K=10$, the tracking performance of the algorithm increases, approaching the centralized performance.

\begin{figure}[ht]
	\centering
\includegraphics[width=3in]{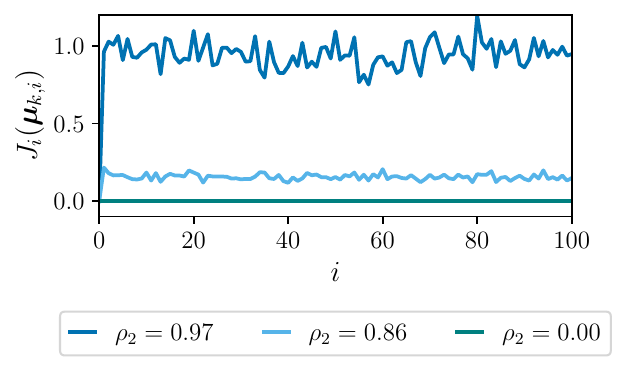}
\caption{Evolution of the risk over time, computed with 1000 Monte Carlo experiments for three different network topologies.}
		\label{fig:j}
\end{figure}

In Fig.~\ref{fig:j}, we observe the evolution of the risk $J_i(\bm{\mu}_{k,i})$ over time for three different graph topologies: $i)$ a very sparse topology with $\rho_2=0.97$, $ii)$ the topology seen in Fig.~\ref{fig:net} with $\rho_2=0.86$, and $iii)$ a fully connected network with $\rho_2=0$. The risk was approximated by averaging 1000 Monte Carlo simulations with the choice of \( \gamma = K \). Fig.~\ref{fig:j} shows that as the second larger eigenvalue $\rho_2$ decreases, the risk approaches zero. This observation is expected in view of Theorem \ref{th:kl_without_net_assumption}.

Having observed that the filter performance increases with step-size chosen around the number of agents, i.e., \( \gamma \to K \), and more graph connectivity, i.e. \( \rho_2 \to 0 \), we finally compare the effect of the network size on the risk function. In Table \ref{tab:num_agents_risks}, the average risk function over the network with respect to different numbers of agents can be found. For all cases, we set \( \gamma = K \). It is hard to get network configurations with exactly the same \( \rho_2 \)'s for different sizes. Therefore, for a fair comparison, we increase the connectivity of the larger graphs, i.e., higher \( K \)'s are associated with smaller \( \rho_2 \)'s. Despite this fact, the average risk increases with increasing number of agents, which supports Theorem \ref{th:kl_without_net_assumption}. This means that the deviation from the optimal centralized algorithm is higher for larger networks, for fixed graph mixing rates.

\newcommand{\ra}[1]{\renewcommand{\arraystretch}{#1}}

\newcolumntype{C}{ >{\centering\arraybackslash} m{2cm} }
\newcolumntype{D}{ >{\centering\arraybackslash} m{2.5cm} }

\begin{table}
\caption{Number of agents and average risks over networks}
\vspace{1em}
\small
\ra{1.5}
\begin{tabular}{  D D D }
  \( K \) &\( \frac{1}{K}\sum_{k=1}^K J_{\infty} (\bmu_{k,\infty}) \)&\( \rho_2 \)\\
 \cline{1-3}
 10   & 0.53&  0.86 \\
 
 20    & 0.83&   0.83\\
 
  30   & 1.17 &   0.81\\
  
  40   & 1.77&  0.80 \\
 
 70    & 2.69&   0.77\\
 
\end{tabular}
\label{tab:num_agents_risks}
\end{table}

\bibliographystyle{IEEEbib}
\bibliography{refs}

\begin{thebibliography}{10}

\bibitem{olfati_2007}
R.~Olfati-Saber,
\newblock ``Distributed {Kalman} filtering for sensor networks,''
\newblock in {\em Proc. IEEE Conference on Decision and Control}, New Orleans,
  LA, USA, 2007, pp. 5492--5498.

\bibitem{khan_2008}
U.~A. Khan and J.~M.~F. Moura,
\newblock ``Distributing the {Kalman} filter for large-scale systems,''
\newblock {\em IEEE Transactions on Signal Processing}, vol. 56, no. 10, pp.
  4919--4935, 2008.

\bibitem{cattivelli_2010}
F.~S. Cattivelli and A.~H. Sayed,
\newblock ``Diffusion strategies for distributed {Kalman} filtering and
  smoothing,''
\newblock {\em IEEE Transactions on Automatic Control}, vol. 55, no. 9, pp.
  2069--2084, 2010.

\bibitem{talebi2021}
S.~P. Talebi, S.~Werner, V.~Gupta, and Y.-F. Huang,
\newblock ``On stability and convergence of distributed filters,''
\newblock {\em IEEE Signal Processing Letters}, vol. 28, pp. 494--498, 2021.

\bibitem{hlinka_2012}
O.~Hlinka, O.~Slučiak, F.~Hlawatsch, P.~M. Djurić, and M.~Rupp,
\newblock ``Likelihood consensus and its application to distributed particle
  filtering,''
\newblock {\em IEEE Transactions on Signal Processing}, vol. 60, no. 8, pp.
  4334--4349, 2012.

\bibitem{battistelli_2014}
G.~Battistelli and L.~Chisci,
\newblock ``{Kullback–Leibler} average, consensus on probability densities,
  and distributed state estimation with guaranteed stability,''
\newblock {\em Automatica}, vol. 50, no. 3, pp. 707--718, 2014.

\bibitem{dedecius_2017}
K.~Dedecius and P.~M. Djurić,
\newblock ``Sequential estimation and diffusion of information over networks: A
  {Bayesian} approach with exponential family of distributions,''
\newblock {\em IEEE Transactions on Signal Processing}, vol. 65, no. 7, pp.
  1795--1809, 2017.

\bibitem{bandyopadhyay_2018}
S.~Bandyopadhyay and S.~Chung,
\newblock ``Distributed {Bayesian} filtering using logarithmic opinion pool for
  dynamic sensor networks,''
\newblock {\em Automatica}, vol. 97, pp. 7--17, 2018.

\bibitem{krishnamurthy_2016}
V.~Krishnamurthy,
\newblock {\em Partially Observed Markov Decision Processes: From Filtering to
  Controlled Sensing},
\newblock Cambridge University Press, 2016.

\bibitem{sayed_2018}
A.~H. Sayed, P.~M. Djurić, and F.~Hlawatsch,
\newblock ``Distributed {Kalman} and particle filtering,''
\newblock in {\em Cooperative and Graph Signal Processing}, P.~M. Djurić and
  C.~Richard, Eds., pp. 169--207. Academic Press, 2018.

\bibitem{papa2019}
G.~Papa, R.~Repp, F.~Meyer, P.~Braca, and F.~Hlawatsch,
\newblock ``Distributed {Bernoulli} filtering using likelihood consensus,''
\newblock {\em IEEE Transactions on Signal and Information Processing over
  Networks}, vol. 5, no. 2, pp. 218--233, 2019.

\bibitem{acemoglu_2011}
D.~Acemoglu, M.~A. Dahleh, I.~Lobel, and A.~Ozdaglar,
\newblock ``Bayesian learning in social networks,''
\newblock {\em The Review of Economic Studies}, vol. 78, no. 4, pp. 1201--1236,
  2011.

\bibitem{krishnamurthy_2013}
V.~Krishnamurthy and H.~V. Poor,
\newblock ``Social learning and {Bayesian} games in multiagent signal
  processing: how do local and global decision makers interact?,''
\newblock {\em IEEE Signal Processing Magazine}, vol. 30, no. 3, pp. 43--57,
  2013.

\bibitem{jadbabaie_2012}
A.~Jadbabaie, P.~Molavi, A.~Sandroni, and A.~Tahbaz-Salehi,
\newblock ``Non-{Bayesian} social learning,''
\newblock {\em Games and Economic Behavior}, vol. 76, no. 1, pp. 210--225,
  2012.

\bibitem{zhao_2012}
X.~Zhao and A.~H. Sayed,
\newblock ``Learning over social networks via diffusion adaptation,''
\newblock in {\em Asilomar Conference on Signals, Systems and Computers}, 2012,
  pp. 709--713.

\bibitem{nedic_2017}
A.~Nedić, A.~Olshevsky, and C.~A. Uribe,
\newblock ``Fast convergence rates for distributed non-{Bayesian} learning,''
\newblock {\em IEEE Transactions on Automatic Control}, vol. 62, no. 11, pp.
  5538--5553, 2017.

\bibitem{lalitha_2018}
A.~Lalitha, T.~Javidi, and A.~D. Sarwate,
\newblock ``Social learning and distributed hypothesis testing,''
\newblock {\em IEEE Transactions on Information Theory}, vol. 64, no. 9, pp.
  6161--6179, 2018.

\bibitem{luengo2018}
D.~Luengo, L.~Martino, V.~Elvira, and M.~Bugallo,
\newblock ``Efficient linear fusion of partial estimators,''
\newblock {\em Digital Signal Processing}, vol. 78, pp. 265--283, July 2018.

\bibitem{matta_2020}
V.~Matta, V.~Bordignon, A.~Santos, and A.~H. Sayed,
\newblock ``Interplay between topology and social learning over weak graphs,''
\newblock {\em IEEE Open Journal of Signal Processing}, vol. 1, pp. 99--119,
  2020.

\bibitem{ntemos_2021}
K.~Ntemos, V.~Bordignon, S.~Vlaski, and A.~H. Sayed,
\newblock ``Deception in social learning,''
\newblock {\em arXiv preprint arXiv:2103.14729}, 2021.

\bibitem{acemoglu_2008}
D.~Acemoglu, A.~Nedic, and A.~Ozdaglar,
\newblock ``Convergence of rule-of-thumb learning rules in social networks,''
\newblock in {\em IEEE Conference on Decision and Control}, 2008, pp.
  1714--1720.

\bibitem{frongillo_2011}
R.~M. Frongillo, G.~Schoenebeck, and O.~Tamuz,
\newblock ``Social learning in a changing world,''
\newblock in {\em Internet and Network Economics}. 2011, pp. 146--157, Springer
  Berlin Heidelberg.

\bibitem{shahrampour_2013}
S.~Shahrampour, S.~Rakhlin, and A.~Jadbabaie,
\newblock ``Online learning of dynamic parameters in social networks,''
\newblock in {\em Advances in Neural Information Processing Systems}, 2013,
  vol.~26.

\bibitem{bordignon_2021}
V.~Bordignon, V.~Matta, and A.~H. Sayed,
\newblock ``Adaptive social learning,''
\newblock {\em IEEE Transactions on Information Theory}, vol. 67, no. 9, pp.
  6053--6081, 2021.

\bibitem{sayed_2014}
A.~H. Sayed,
\newblock ``{Adaptation, learning, and optimization over networks},''
\newblock {\em Foundations and Trends in Machine Learning}, vol. 7, no. 4-5,
  pp. 311--801, July 2014.

\bibitem{dobrushin_1956}
R.~L. Dobrushin,
\newblock ``Central limit theorem for nonstationary {Markov} chains. {I},''
\newblock {\em Theory of Probability \& Its Applications}, vol. 1, no. 1, pp.
  65--80, 1956.

\bibitem{polyanskiy_2017}
Y.~Polyanskiy and Y.~Wu,
\newblock ``Strong data-processing inequalities for channels and {Bayesian}
  networks,''
\newblock in {\em Convexity and Concentration}, New York, NY, 2017, pp.
  211--249, Springer New York.

\bibitem{metropolis_1953}
N.~{Metropolis}, A.~W. {Rosenbluth}, M.~N. {Rosenbluth}, A.~H. {Teller}, and
  E.~{Teller},
\newblock ``Equation of state calculations by fast computing machines,''
\newblock {\em The Journal of Chemical Physics}, vol. 21, no. 6, pp.
  1087--1092, Jun 1953.

\end{thebibliography}

\end{document}